# Precise wavefront characterization of X-ray optical elements using a laboratory source

J. Lukas Dresselhaus[a], Holger Fleckenstein[b], Martin Domaracký[b], Mauro Prasciolu[b], Nikolay Ivanov[b], Jerome Carnis[b], Kevin T. Murray[b], Andrew J. Morgan[c], Henry N. Chapman[a,b,d], and Saša Bajt[a,b]

[a]The Hamburg Centre for Ultrafast Imaging, Luruper Chaussee 149, 22761 Hamburg, Germany

[b]Center for Free-Electron Laser Science CFEL, Deutsches Elektronen-Synchrotron DESY, Notkestr. 85, 22607 Hamburg, Germany

[c]School of Physics, University of Melbourne, Parkville, Victoria 3010, Australia,

[d]Department of Physics, Universität Hamburg, Luruper Chaussee 149, 22761 Hamburg, Germany

## Abstract

Improvements in X-ray optics critically depend on the measurement of their optical performance. The knowledge of wavefront aberrations, for example, can be used to improve the fabrication of optical elements or to design phase correctors to compensate for these errors. Nowadays, the characterization of such optics is made using intense X-ray sources such as synchrotrons. However, the limited access to these facilities can substantially slow down the development process. Improvements in the brightness of lab-based X-ray micro-sources in combination with the development of new metrology methods, and in particular ptychographic X-ray speckle tracking, enable characterization of X-ray optics in the lab with a precision and sensitivity not possible before. Here, we present a laboratory set-up that utilizes a commercially available X-ray source and can be used to characterize different types of X-ray optics. The set-up is used in our laboratory on a routine basis to characterize multilayer Laue lenses of high numerical aperture and other optical elements. This typically includes measurements of the wavefront distortions, optimum operating photon energy and focal length of the lens. To check the sensitivity and accuracy of this laboratory set-up we compared the results to those obtained at the synchrotron and saw no significant difference. To illustrate the feedback of measurements on performance, we demonstrated the correction of the phase errors of a particular multilayer Laue lens using a 3D printed compound refractive phase plate.

## 1. Introduction

Optics with low aberrations are prerequisite for high resolution X-ray microscopy. As in all precision optical instrumentation, the measurement of the optical wavefront is critical for the development of X-ray optics. Such data is instrumental to pinpoint the origin of errors and to make changes in the fabrication process to reduce aberrations. It is well known that the quality of the optical element that you can make critically depends on how accurately you can measure it. Access to high brightness X-ray sources, such as synchrotrons and X-ray free electron lasers (XFELs), is highly competitive and should be pursued only when really needed. Laboratory X-ray sources are attractive because of their lower cost of use and ease of access. Here we present a set-up that is used on a routine basis in our laboratory to characterize various X-ray optical components. In the following we place an emphasis on wavefront measurements of multilayer Laue lenses (MLLs) [19–21] although the same set-up can characterize compound refractive lenses, zone plates, or can measure the efficiency of X-ray gratings, for example. The primary development in our laboratory is of multilayer-based optics, consisting of many tens of thousands of layers, with a total lens height that can exceed 100 μm, and with individual layers that can be thinner than 1 nm (see Fig. 1).

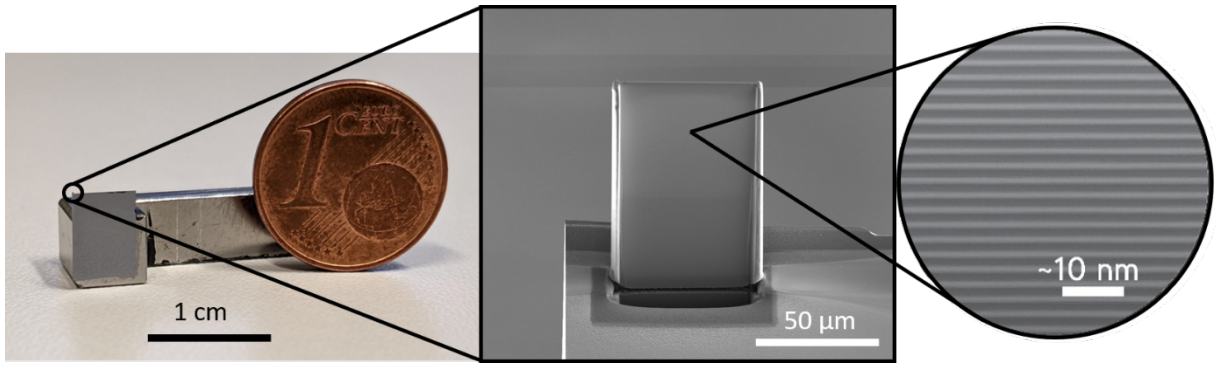

Fig 1: Left: Photograph of a lens holder with a 100 μm thick Si wafer glued on it. An MLL is mounted on the top left corner of the wafer. Middle: Scanning electron microscopy (SEM) image of the MLL consisting of ~30 000 individual layers. Right: Transmission electron microscopy (TEM) image showing several individual layers in cross section.

The deposition of such thick multilayers can take several days. Systematic changes or instabilities of the sputter rate of the target materials lead to a systematic error in the layer thickness and hence an error in the layer placement in the lens, leading to a phase error. We wish to discover and understand these errors and their causes, so that they can be eliminated or compensated.

Various wavefront characterization methods, including grating-based [7,14,22,23] and speckle-based [24–26] wavefront sensing approaches have been pursued in the past. Some of these have been implemented using laboratory X-ray sources [8]. We use ptychographic X-ray speckle tracking (PXST) [27], which combines X-ray speckle-based methods with a ptychographic approach to retrieve the stationary wavefront and the phase induced by a scanned object. As with other related speckle-tracking methods [28,29], phases are obtained non-interferometrically which avoids high requirements of temporal and spatial coherence of the X-ray source. Wavefront sensing based on speckle tracking was successfully demonstrated using laboratory-based X-ray sources [28,29]. The theory, experimental applications and the source code for PXST have been described extensively elsewhere [27,30,31]. Our laboratory-based X-ray measurement station combines a commercial high brightness X-ray source (Sigray, Inc.) with our in-house developed instrumentation and controls, as described here. After we introduce the experimental set-up, a short introduction of MLLs and their preparation is given, followed by the wavefront sensing method (Sec. 5). We then compare the results measured with this laboratory set-up with those obtained at the synchrotron, in Sec. 6. Based on the measured wavefront we designed and prepared a phase corrector for a particular MLL, which was then applied to obtain an improved wavefront (Sec. 7).

## 2. Experimental set-up

The laboratory set-up utilizes a commercially available X-ray micro-source, Sigray XCITE Illumination Beam system (Sigray Inc., USA). This source contains a fine array anode structured target (FAAST™) embedded in a diamond substrate, which allows for efficient thermal dissipation. Hence, even materials with low thermal conductivity can be used as X-ray emitters [5]. The choice and the number of anode materials in the FAAST™ can be customized. Our source includes Cu, W, Mo and Rh targets, all positioned on one anode. Switching between different anode materials is quick and accomplished with simple software commands. The X-ray source size is $15 \pm 1$ μm x $12 \pm 1$ μm in the horizontal and vertical directions, respectively, as estimated from a knife edge scan. The maximum source power depends on the anode material: 30 W for Cu, 40 W for W and Mo, and 50 W for Rh. However, all the data presented here were obtained with the Mo anode (characteristic line: Mo K$\alpha$ at 17.48 keV) [32]. According to the manufacturer, the brightness of the Mo anode is $1.7 \times 10^{10}$ photons $s^{-1}\text{mm}^{-2}\text{mrad}^{-2}$.

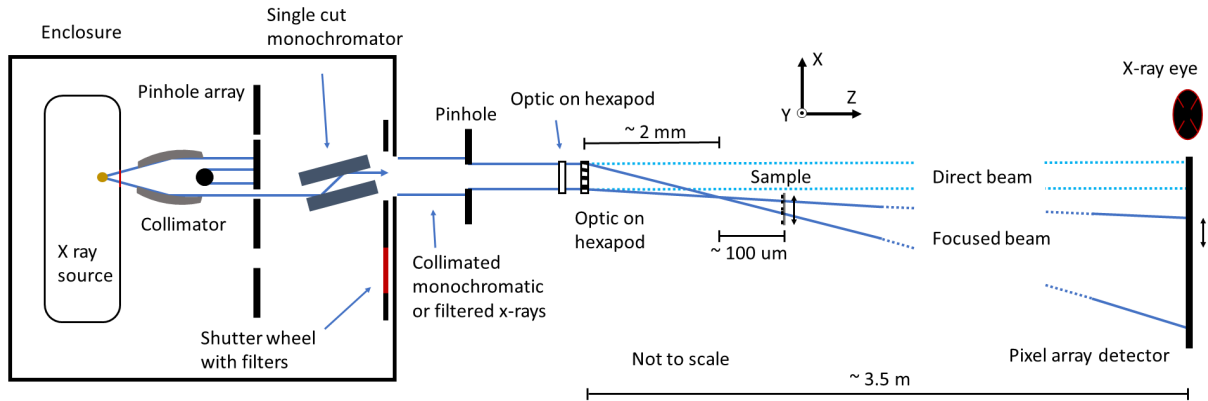

Fig 2: Schematic of the X-ray set-up: A lead box with the X-ray source, collimator, pinhole array, monochromator and a wheel with filters (which is also used as a shutter) is located inside a larger enclosure (not shown) sitting on a 5 m long optical table. The X-ray beam that emerges from the lead enclosure passes through a 1 mm diameter pinhole which blocks scattered and stray radiation. Two 6-axis hexapods are used to hold and align X-ray optics or other components. A sample sits on a stage with *x, y, z* motion and a rotation adjustment in the *x-y*-plane. Two detectors, an X-ray eye and a pixel array detector, are mounted on a common *x, y* stage. The detectors can be manually moved along the *z* direction (along the beam path), guided by a central rail system.

Figure 2 shows a schematic of the experimental set-up and indicates our right-handed coordinate system with *z* in the beam direction. The entire set-up is mounted on a 5 m long air floating optical table which is surrounded by a radiation protection enclosure. The enclosure has two lead-glass windows to see the set-up without breaking the interlocks. The X-ray source itself is inside an additional enclosure, a lead box with walls consisting of Pb plates sandwiched between Al plates. Inside the box, the X-rays emitted from the source are first collimated achromatically with a Pt-coated paraboloidal mirror [33] with 44 mm working distance. This collimator is 25 mm long with 0.56 mm entrance and 0.70 mm exit diameters, respectively. It was designed and fabricated by Sigray Inc. Switching to a different anode material requires only minor re-alignment of the collimator. The central part of the beam, which passes through the optic without being reflected, is blocked at the exit with a Pt beam stop. The effective numerical aperture (NA) of the optic is 0.0128. The size of the collimated beam is then reduced with a pinhole, which can be selected from a set of pinholes with diameters of 150 μm, 75 μm and 45 μm. Optionally, the collimated beam can be monochromatized using a Si (111) channel-cut monochromator located about 17 cm from the source. The monochromator, which was designed and fabricated at DESY, can accept and monochromatize X-rays between 7 and 21 keV. The expected angular acceptance of the double Bragg reflection for 17.5 keV is 15 μrad, which is equivalent to an energy bandwidth of $1.3 \times 10^{-4}$.

The monochromator shifts the X-ray beam by $5 \cos \theta$ (mm), where $\theta$ denotes the Bragg angle. Hence, when the beam is monochromatized, all other components downstream must be shifted accordingly. However, for many applications it is sufficient to use a foil filter instead of a monochromator. This results in a beam with a broader energy bandwidth but with noticeably higher flux. Using appropriate filter materials and filter thickness, one can absorb the K$\beta$ line without significantly absorbing the K$\alpha$ line. For the Cu source we employ a 25 μm thick Ni filter and for Mo a 50 μm thick Nb filter. A 25 μm thick Cu filter can be used for the W anode to suppress the *Lβ* line. These filters are mounted on a shutter and filter wheel as shown in Fig. 2. Outside the source enclosure, an aperture of 1 mm diameter is used to reduce stray X-ray light.

The X-ray beam that emerges from the source enclosure is used for a variety of experiments that can be arranged on the large optical table. The wavefront of high-NA optical elements such as MLLs can be characterized by our method of PXST described in Sec. 5. For this, the optic is made to focus the collimated beam and a test structure is scanned through the diverging beam. Projection images of that structure are recorded on a pixel-array detector. Two SmarAct (SmarAct, GmbH, Germany) hexapods are available for positioning pairs of MLLs or a single MLL with a phase plate corrector. The resolution of length adjustments of the hexapod motors is about 1 nm, providing a similar positioning resolution in each of three orthogonal directions and an angular resolution of 1 μrad in three orthogonal directions. The angular range of the hexapod tilt is 16°. The hexapods are mounted on a large frame. One hexapod sits on the bottom of the frame while the other hangs from the top. Optical elements are mounted on metal holders (see Fig. 1) that can slide into magnetic receptacles on the hexapods. This makes the exchange of optics and samples very easy and reproducible (within a few tens of micrometers). The sample is mounted on a scanning stage, which consists of a stack of three SmarAct translation stages with 10 nm positioning precision and one SmarAct rotational stage that can rotate the sample in *x-y*-plane with 0.4 μrad precision. The sample is usually fixed on a pin as shown in Fig. 3, which is mounted on a magnetic base attached to the sample scanner. The distance between the last hexapod and the sample is adjustable and can be as large as 3 m, only limited by the size of the optical table. Due to the precise and controlled movement of the sample stage and hexapods, it is safe to position two elements within a distance of only a few micrometers from each other.

All motors, with the exception of the motors moving the Sigray X-ray source, are controlled by our in-house developed software (Kamzik3) [34]. This modular control framework is based on Python3. It uses ZeroMQ to exchange messages between the server and clients and Qt5 for its graphical interface. Kamzik3 also provides tools for logging, visualizing and evaluating experimental data. With a built-in macro-server one can create and execute custom macros and scans. These scans include 1D and 2D scans with and without acquiring detector frames. The experimental set-up is defined in the configuration file written in YAML, which is human-readable and the framework is available under the GNU General public license [34].

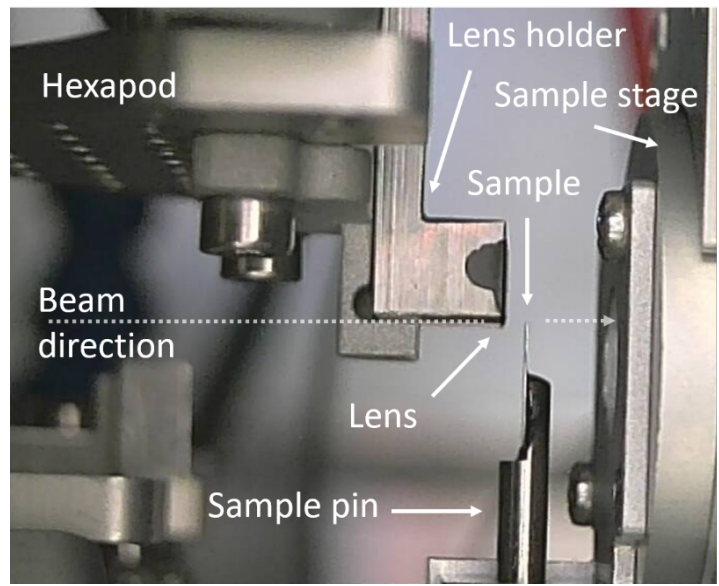

Fig 3: The MLL lens and the sample on their stages in the laboratory set-up. The lens sits on the lower edge of the holder closest to the sample pin and the sample is positioned on the top edge of the thin silicon chip, mounted on a metal pin. The beam propagates from left to right. The detector (not shown) is far to the right. The part labelled "Sample stage" actuates the rotation of the sample and has a hole to allow the beam to pass.

The experimental set-up utilizes two detectors mounted on a common movable platform: a scintillator imaged onto a PCO 4.2LT CCD (Excelitas PCO GmbH, Germany) camera, referred to as an X-ray eye, and a LAMBDA 750k (X-Spectrum GmbH, Germany) photon counting pixel array detector. The effective pixel size of the X-ray eye is 0.65 µm, but the resolution of the 10× objective used to image the scintillator limits the resolution of this device to about 2 µm. The silicon sensor of the LAMBDA detector (Medipix3) is comprised of 516×1556 square pixels, each with a size of 55 µm. The PCO camera is used for initial alignment while the LAMBDA detector is used for the wavefront sensing measurements. The largest distance between the sample and the detectors is 3.6 m, limited by the size of the enclosure. The positioning of the detector platform is motorized in the *x* and *y* directions and can be pushed manually along a rail in *z*. The detector images are stored in *nxs* files, which are structured like *h5* files. Four miniature cameras, Watec WAT-240E (Watec GmbH, Germany) are used to view the optical elements (lenses) and the sample from various directions. This motorized set-up with view cameras allows for safe remote operation at a fixed sample to detector distance once an optic and a sample are installed.

## 3. Multilayer Laue lenses and their preparation

The X-ray optics discussed here are based on multilayers fabricated in our laboratory using magnetron sputtering deposition [35]. Each layer period consists of a pair of layers of two materials, alternately deposited on a substrate by moving it in and out of the plasmas of two magnetrons with different target materials. The amount of time the substrate spends in each plasma determines the layer thickness. In MLLs, the period thickness $d_n$ varies inversely with the distance $r_n$ of the layers from the optical axis, to ensure that incident rays are diffracted to a common focus, approximately as

$$d_n = \lambda f / r_n, \tag{1}$$

for a focal length $f$ and wavelength $\lambda$. An MLL that focuses in two dimensions must be deposited on a cylindrical substrate (such as a wire) starting with the thickest and ending with the thinnest layer. However, when depositing layers in this order it is challenging to maintain the smoothness of layers since the initial, thicker layers tend to be polycrystalline and rougher. Deposition on a flat substrate, starting with the thinnest and ending with the thickest periods, solves this problem, and is more efficient and easier to control. This and other challenges associated with MLL preparation were described in a review paper [36]. The MLL cut from a flat substrate focuses X-rays only in one direction. Two of them, placed together in orthogonal orientations, are required to achieve a 2D focus [37].

The multilayers in this study consist of tungsten carbide (WC) and silicon carbide (SiC) or tungsten (W) and silicon carbide [19,20]. To achieve high diffraction efficiency, each layer of an MLL must be tilted such as to obey Bragg's law (see Fig. 4). Due to the variation in period (Eqn. 1), this tilt must vary linearly with position $r_n$. We use a simple method to achieve such wedging using a straight mask [35]. The mask, which is attached above the substrate and moves with it, casts a shadow, which automatically results in a thickness gradient of the layers [38]. After the deposition, the thickness profile is measured with the help of a focused ion beam (FIB, FEI Helios NanoLab, Thermo Fisher Scientific, USA). The position where the layer gradient matches the MLL design can be determined within a few tens of micrometers.

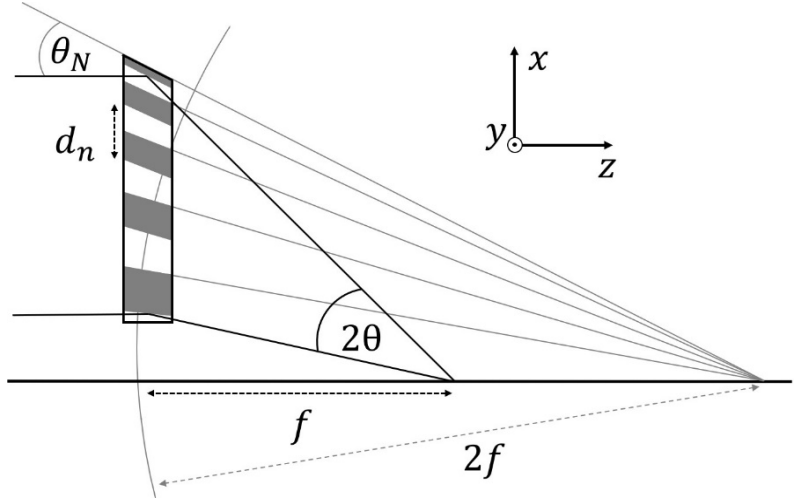

Fig 4: Schematic of the operating principle of an MLL. The period thickness reduces with distance from the optical axis according to the zone plate condition and each layer is tilted to obey Bragg's diffraction condition. This is achieved if the layers are normal to a circle with radius $2f$. The lens pupil, in this example, is off axis as the layers do not extend all the way to the optical axis.

High efficiency lenses require the correct optical thickness. The optimum optical thickness of a lens can be calculated using dynamic diffraction theory and depends on the properties and the layer structure of the multilayer as well as on the energy at which the lens is to be used [39]. The cutting of MLLs is also performed with the FIB and can take several days for a 100 μm thick multilayer. This process can be substantially shortened with a Xe+ plasma FIB (PFIB) which is up to 60 times faster than cutting with a "standard" Ga+ FIB. The MLL slice is then glued to the tip of a micromanipulator, extracted, transferred and affixed to an edge of a silicon wafer which itself is attached to a lens holder (see Fig. 1). From a single multilayer deposition one can extract many identical lenses, which are optimized for the same energy and have the same focal length. However, from a single multilayer deposition one can also prepare many MLLs optimized for different energies or extract MLLs that work at the same energy but have different focal lengths. The latter can be achieved by using two masks at different mask-to-substrate distances [38].

## 4. Optimum energy determination

To achieve high diffraction efficiency all layers in the lens have to be tilted according to their respective Bragg angle. This can only be realized for a single photon energy (wavelength). If the incident X-ray photon energy differs from this, only a fraction of the lens will diffract the X-rays. The range of scattering angles ($2\theta$) changes linearly with the tilt $\omega$ of the lens. The slope depends on the ratio between the wavelength of the photons used to probe the lens $\lambda_p$ and the ideal operating wavelength of the lens $\lambda_m$:

$$\omega = \frac{1}{2}\left(1 - \frac{\lambda_m}{\lambda_p}\right) 2\theta \qquad (2)$$

By tilting the lens along $\omega$ and recording the intensity distribution along $2\theta$, the optimum operating wavelength of a lens $\lambda_m$ can be determined [19]. Since one lens focuses only in one direction, the diffracted beam also converges only in one direction, to a line focus which diverges from there to be measured on a pixel-array detector (LAMBDA detector, see Fig. 2).

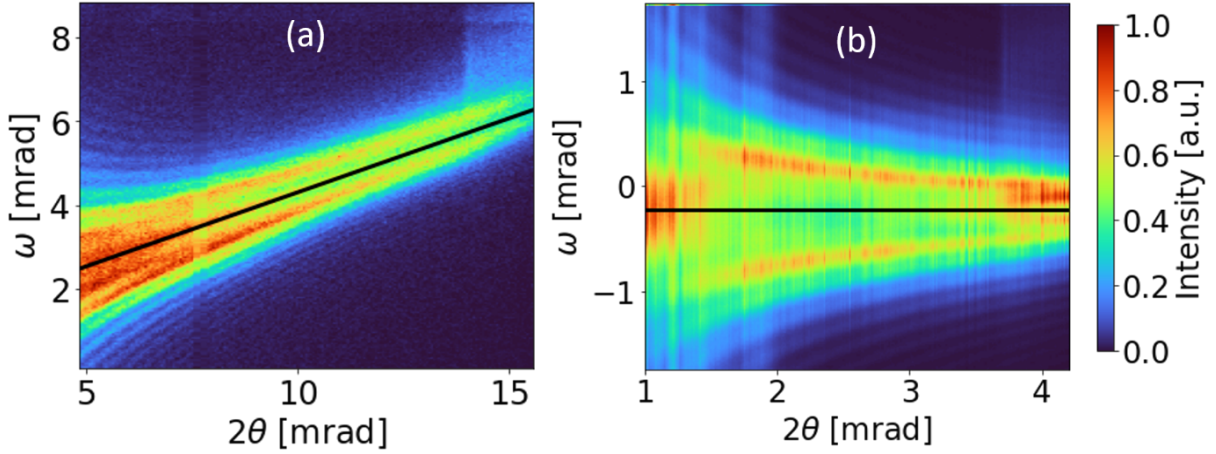

Fig 5: Plots displaying MLL's angle tilt ($\omega$) vs. $2\theta$ (diffraction line) for a MLL designed for 60 keV. The black line indicates the slope from which the ideal operating energy was calculated. (a) The measurement was performed with a laboratory set-up using 17.5 keV. (b) The same lens measured with 60 keV synchrotron radiation (P07 beamline, PETRA III). In this case the probing energy matches the optimum energy of the MLL.

The in-plane orientation of the lens is adjusted such that the diffracted line is parallel to the pixel rows of the LAMBDA detector and the signal is integrated along pixel columns. This results in a one-dimensional map of diffraction efficiency across the lens pupil (i.e. as a function of $2\theta$ at a particular lens tilt $\omega$). By stacking those, a two-dimensional map is obtained such as shown in Fig. 5.

The location of the maximum efficiency at each $2\theta$ gives a plot (black lines) showing the relationship between $\omega$ and $2\theta$, from which the operating wavelength of the lens $\lambda_m$ can be derived using Eqn. 2. For the particular sign convention of $\omega$ used here (positive rotates the thinner layers towards the focus), a positive gradient indicates that the optimum photon energy of the lens is higher than the probe energy, and vice versa. Figure 5 (a) shows a lens made for operating at 60 keV, measured using 17.5 keV. Examples of the operating photon energies of some lenses are provided in Table 1.

The probing wavelength $\lambda_p$ does not have to be close to $\lambda_m$ for a precise wavelength (energy) determination. In fact, a single probing wavelength $\lambda_p$ is sufficient to test any lens. This is convenient since it enables us to prepare and measure the energies of lenses needed for experiments at other facilities. The measurement made with 17.5 keV and shown in Fig. 5 (a) was confirmed by a measurement at the operating energy of 60 keV, made at the P07 beamline of PETRA III.

| Name | Energy [keV] | 2NA @ 17.5 keV | Focal length @ design energy [mm] | Height [µm] | Optical thickness [µm] |
|---|---|---|---|---|---|
| LC1 | 60 | 0.01 | 15 | 52 | 35 |
| LS1 | 17.5 | 0.03 | 1.15 | 34 | 9.3 |
| LS2 | 17.5 | 0.03 | 1.25 | 35 | 9.6 |
| LX1 | 9.3 | 0.008 | 8 | 89 | 4.9 |

Table 1: Parameters of some lenses characterized at 17.5 keV (using a Mo Kα source).

# 5. Determination of MLL wavefront aberrations

To characterize the quality of lenses we use PXST, a wavefront sensing method well suited for the laboratory set-up [27]. Despite the name, it does not require the formation of speckles, but tracks intensity features in projection holograms of a transmitting object placed in the diverging wavefield. The object is stepped across the beam so that there is a high degree of overlap of the illuminated region and such that any particular feature is visible in numerous positions across the lens pupil, giving a high redundancy of measurements. Differences of the measured positions of features from those expected (given the magnified step size) are a result of phase gradients that differ from the idea spherical or conical wave. The phase gradient is reconstructed iteratively by alternatively updating estimates of it and an aberration-free image of the sample. Once the results converge, the wavefront can be calculated by integrating the phase gradient. More details, including the software suite and examples, are provided in references [27,30,31].

Our implementation of the PXST algorithm can be applied to characterize the two-dimensional wavefront, formed from an orthogonal pair of MLLs for example, or the one-dimensional wavefront of a single lens. Of course, to measure the wavefront of the full lens, the probe wavelength must match the operating wavelength of the lens. In the laboratory set-up we usually characterize one lens at a time. A single MLL acts as a cylindrical lens and hence creates a line focus from which a cylindrical wave diverges. This wave is characterized using a 11.6 µm tall aperiodic multilayer structure that resembles a "barcode". This particular structure consists of 18 SiC and 18 W layers. This choice of materials ensures projection images of high contrast, which is beneficial for quick and easy alignment as well as for reconstruction of the wavefront with high precision. While all the SiC layers are 180 nm thick, the thicknesses of the W layers range between 200 nm and 1000 nm following a random pattern.

To measure the wavefront of an MLL, the barcode sample is oriented such that its layers are parallel to the line focus of the lens, e.g. orthogonal to the direction of the diverging wave. This is accomplished using the in-plane rotation adjustment of the sample stage, to which the sample mounting pin is attached. Without this adjustment, the one-dimensional projection image will be blurred and of lower contrast. The lens, mounted on a hexapod, is initially aligned using the X-ray eye detector. At the Bragg condition, beam extinction causes the transmission image of the MLL to appear much darker than the semi-transparent Si wafer substrate. Precise alignment is finalized by tilting the lens to maximize the intensity of the divergent beam on the LAMBDA detector. If the operating wavelength does not match the probe wavelength, only a fraction of the expected divergent beam is visible for any given lens tilt. Since the lens pupil of an MLL is off axis, the focused beam is clearly and distinctly separated from the on-axis zero-order (unfocussed) beam.

The barcode sample is placed slightly downstream of the focus to create a magnified projection image that is seen as a modulation of the one-dimensional divergent beam on the detector. The magnification factor $M$ is given by the ratio between the distance from the focus to the detector $D$ and the defocus distance of the sample $\Delta f$, $M = D/\Delta f$. While the detector distance in our set-up can be as large as 3.6 m, it is also limited by the distance at which the divergent line still fits on the detector chip. Placing the detector closer to the sample reduces the required measurement time but this is at the expense of reduced measurement sensitivity. Measurements at higher magnifications (shorter defocus) are more sensitive to the wavefront aberrations. However, if the sample is too close to the focus, sample features cannot be tracked anymore. This distance depends on the magnitude of the wavefront aberrations. Typically, the detector is placed 2 m downstream from the sample and the defocus distance is around 100 µm, for a magnification of 20000. With a pixel size of 55 µm on the LAMBDA detector, the resulting effective pixel size is 2.75 nm. (The reconstructed image does not achieve this resolution due to the finite angular extent of the source.)

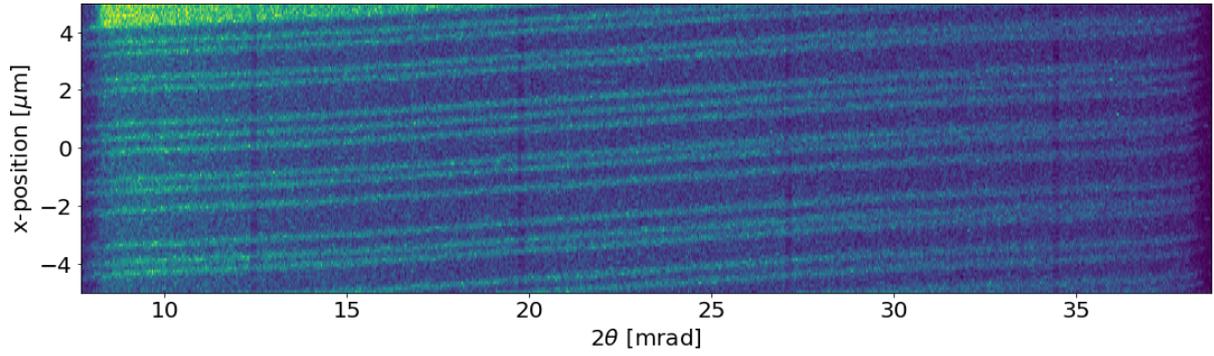

Fig 6: Ptychograph of the barcode sample, measured with lens LS1 at a magnification of 25000, a step size of 100 nm, and a photon energy of 17.5 keV (0.071 nm wavelength) using the laboratory source.

A speckle-tracking measurement scans the sample transversely through the divergent beam with a fixed step size. At each scan point, the measured detector frame is integrated along the pixel direction perpendicular to the divergent line as described in Sec. 4. A near-field ptychograph is created by stacking these integrated one-dimensional images as shown in Fig. 6 for the case of lens LS1 (see Table 1). The choice of the scan step size depends on the magnification and the desired redundancy. Here, the sample was placed 80 μm behind the focus and the detector was 2 m downstream. This gave magnification of 25000. In this case a step size of 100 nm was appropriate. Since our sample is 11.6 μm tall, 100 steps were sufficient to fully cover it. Note that it is not strictly necessary to cover the full sample as long as at least several features can be tracked to determine the wavefront. Reducing the number of steps to 30 and using the same step size (100 nm) resulted in no change of the reconstructed wavefront. For this measurement, we used an exposure time of 30 s per step. Hence, the total scan was accomplished in less than 1 hour.

Since during these measurements the sample is defocused, the barcode features (layers) move across the field of view as the sample is moved. For an aberration free lens, the resulting lines in the ptychograph would be straight (linear). Any deviation from a straight line is a direct result of the phase gradient due to aberrations. Indeed, the trajectory of a feature in the ptychograph maps out the phase gradient: pure defocus gives a straight line, a third-order aberration (coma) gives a parabola, and so on. The PXST calculation, as a generalization of the Hartmann method [40], provides this phase gradient in a consistent manner from all features and usually converges within a few iterations, which takes only a few seconds.

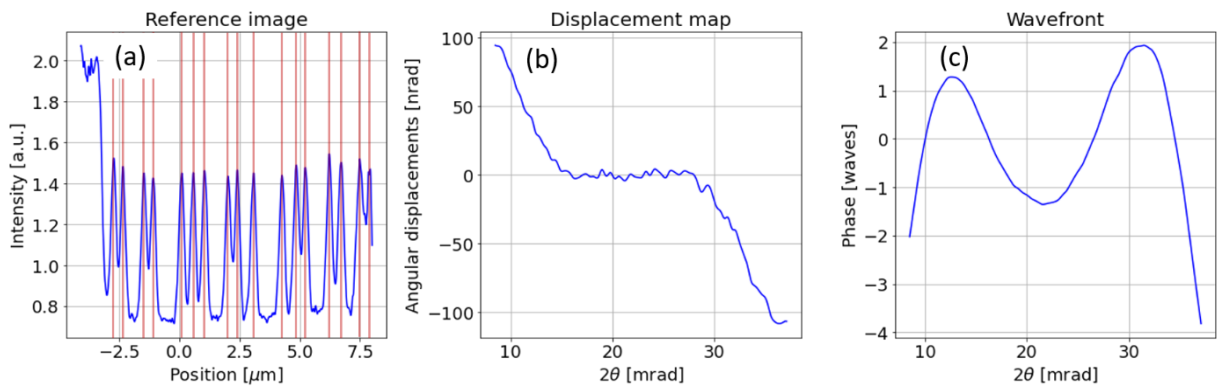

Fig 7: Results of a 1D speckle tracking analysis of the ptychograph measurement shown in Fig 6. (a) The 1D reference image with underlaying actual positions of the SiC layers in red. (b) Displacement map, (c) and wavefront of lens LS1.

Reconstructions of the image of the barcode object and the displacement map representing the phase gradient of the wavefront of lens LS1, obtained using the PXST software, are shown in Fig. 7 (a) and (b), respectively. The image and phase gradient were updated alternately until they converge to a consistent solution. The wavefront of the lens was then calculated by integrating the phase gradient (Fig. 7(c)). From multiple repetitions of the measurement, we estimated the precision of the phase gradient to be about 18 µrad. Although the PXST software was originally developed for use with data collected using synchrotron radiation it works well also on data with lower signal to noise ratio as demonstrated here.

The phase gradient of the lens can be fitted with a third order polynomial while the phase can be described with a fourth order polynomial. This implies a residual quadratic change in the sputter rate of the deposition process [41] that remains after rectifying the linear drift in the fabrication process. This additional correction to the change in sputter rate can be fed back into the fabrication process.

## 6. Comparison with synchrotron measurements

The wavefronts of two individual MLLs (LS1 and LS2) were determined with our lab set-up, as described in Sec. 5, before using them as a pair to form a two-dimensional focus in an experiment at beamline P11 of the PETRA III synchrotron radiation facility (DESY, Hamburg, Germany). The two lenses were cut from a single multilayer, which was deposited using two masks with different distances to the substrate. Hence, lenses of the same energy but slightly different focal lengths and numerical apertures [38] could be cut from the same multilayer—here, the focal lengths were 1.15 mm and 1.25 mm.

To characterize the focus of the lens pair, the wavefront was measured by PXST using a Siemens star object with innermost spokes of 50 nm (see Fig. 8 (a)). The scan consisted of a grid of 20 × 20 steps with a step size of 150 nm at a defocus distance of ~100 µm. The projection images were recorded using an Eiger16M detector (Dectris AG, Switzerland) positioned at a distance of 2.37 m from the focus, providing a magnification of 23700 and an effective pixel size of 3 nm in both directions. The detector consisted of 4096 × 4096 pixels with 75 µm individual pixel size, but the area covered by the lens pupil was about 942 × 904 pixels. The images were acquired with an exposure time of 100 ms.

The 2D phase map (see Fig. 8 (b)) of the lens pair was retrieved using PXST as described above. The phase map is described predominantly by two orthogonal separable functions in the vertical and horizontal directions, which are plotted in Figs. 9 (a) and (b), respectively. In addition, we observed a small amount of 45-degree astigmatism, due to orthogonal misalignment of the two lenses [37] and which is not separable. The plots of Figs. 9 (a) and (b) also show the wavefronts measured individually from each lens with the laboratory set-up. An excellent match of phases of both lenses is observed. Small differences may be due to non-perfect separability of the 2D phase map into individual wavefronts and the precision of the measurement. Figure 9 (c) shows the point spread function (PSF) calculated for the combination of both lenses. The resulting FWHM is 4.3 nm × 4.5 nm.

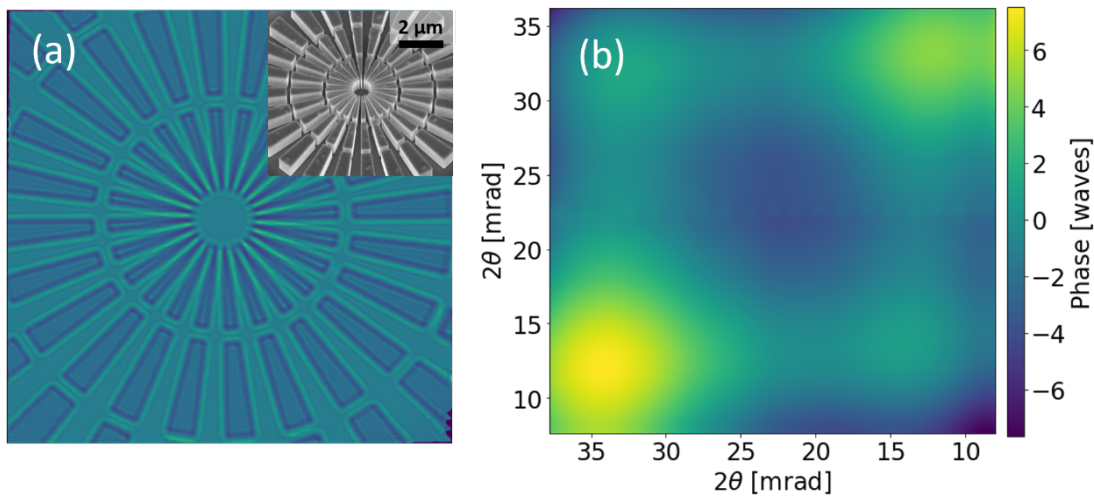

Fig 8: (a) Reconstructed projection hologram of a Siemens star test pattern. The inlet shows an SEM image of the object. (b) 2D phase map of lenses LS1 and LS2 measured at P11 beamline (PETRA III).

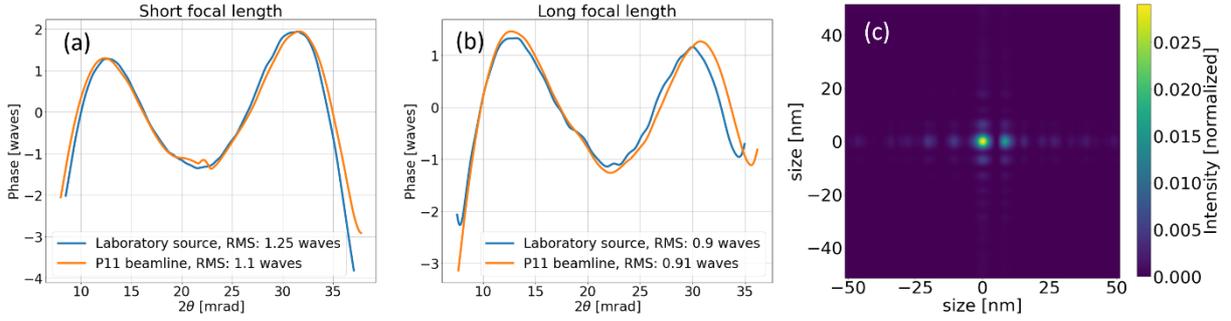

Fig 9: Comparison of the wavefronts obtained with the laboratory (blue) and synchrotron (orange) set-ups for (a) a lens of 1.15 mm focal length and (b) 1.25 mm focal length. The lenses have almost the same NA and optimal photon energy. The laboratory measurements were made on single lenses while the wavefronts obtained at the synchrotron were extracted from a 2D measurement made on a lens pair. (c) The PSF calculated for a combination of both lenses. The intensity is normalized to the maximum obtainable for a perfect wavefront. The FWHM is 4.3 nm × 4.5 nm

## 7. Phase correction

The measurement of wavefront aberrations can help to make improved lenses. Alternatively, if the wavefront aberrations are small enough they can be corrected using a phase plate. In the X-ray regime, phase correcting elements have been successfully implemented to improve wavefronts of compound refractive lenses [42–44] and Kirkpatrick–Baez focusing mirrors [45,46]. The same principle can be used to reduce phase aberrations in MLLs.

The phase profile of a phase plate is designed to conjugates the wavefront of the lens and create a uniform phase when combined with the lens. X-ray phase correctors often utilize refraction and can be manufactured using additive or subtractive approaches [47]. A subtractive approach was demonstrated using laser ablation [48] while additive correctors can be made by 3D printing [42–44]. The advantage of 3D printing is the high degree of control of the printing parameters and the freedom to print almost any arbitrary shape. Two-photon polymerization printing produces structures of high resolution from polymer materials, which have low X-ray absorption and thus are well suited to create X-ray optical elements. The required thickness $t(r)$ of a material of refractive index $n = 1 - \delta$ to correct a wavefront $\phi(r)$ (measured in radian) is given by

$$t(r) = \frac{\lambda}{2\pi\delta} \phi(r). \qquad (3)$$

For hard X-rays such as the 17.5 keV photon energy considered here, the refractive index decrement $\delta$ of materials of low atomic number is of the order of $10^{-6}$ and thus refractive correctors require millimeter thicknesses to induce several waves of phase change. This is comparable to the focal lengths of our lenses. Thus, such a corrector must be placed upstream of the lens. Also, given that the sizes of lens apertures used here are less than 100 μm, corrector structures tend to require large aspect ratios with surfaces that approach grazing incidence to the incoming beam. To contend with this, correctors are divided into $N$ identical elements, each correcting $1/N$ of the wavefront. Such compound refractive correctors [49] reduce the required high aspect ratio and simplify the 3D printing process.

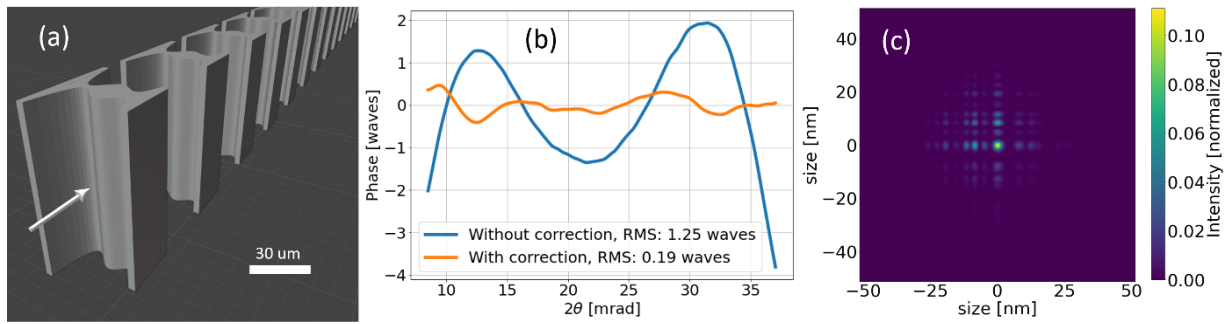

Fig 10: (a) Design of a compound phase plate corrector. The white arrow indicates the beam direction. (b) The measured phase aberration of lens LS1 before (blue) and after correction with the phase plate corrector (orange). (c) The PSF calculated for two orthogonal lenses with the same corrected wavefront given in (b), for comparison with Fig. 9 (c). The intensity is normalized to the maximum obtainable for a perfect wavefront. The FWHM is 2.6 nm × 2.6 nm.

Absorption limits the magnitude of wavefront errors that can be corrected, since large errors would require structures that are too thick to transmit the beam. Even with compound structures thinner than given by that limitation we must take into account how the phase evolves as the wave propagates from the structure to the lens. To account for this, we simulated the phase propagation from the structure to the lens using ray tracing to update the design of the corrector.

A phase corrector for the LS1 lens was prepared using a Nanoscribe (Nanoscribe GmbH, Germany) two-photon polymerization 3D printer [50]. It allows printing of structures of centimeter size with a precision of about 200 nm. The peak-to-valley wavefront error of LS1 is about 5 waves which was corrected using 20 elements with a total length of 1.4 mm. Figure 10 (a) shows the CAD design of the first few elements and their dimensions. Since $n < 1$ for the printed material, a greater thickness of material advances the phase relative to propagation though air. The structure was printed on ITO-coated glass using IP-S as a resist and a 25× magnification objective. The printed structure was developed in PGMEA for 30 minutes and rinsed is isopropanol for 5 minutes. The final step included UV-curing for 30 minutes to harden the structure.

The wavefront corrector was placed on the second hexapod in the laboratory set-up, allowing alignment with six degrees of freedom. Precise alignment was enabled using additional structures that were printed upstream and downstream of the corrector. Alignment with the incident beam direction was achieved when the shadows of these structures overlapped, as seen on the X-ray eye. The corrected wavefront of the LS1 lens is shown in Fig. 10 (b), compared with the uncorrected wavefront. The wavefront aberrations were reduced from 1.25 waves RMS to 0.19 waves RMS. With this correction, the resulting FWHM of the PSF is 2.6 nm (see Fig. 10 (c)).

## 8. Summary

We have developed a laboratory measurement station for the characterization of refractive or diffractive X-ray focusing optics, such as high-NA multilayer Laue lenses, using a high-brightness microfocus source. The system is used to determine the performance and characteristics of optical elements to give rapid feedback in the manufacture of lenses, the development of new designs, and in the preparation of experiments at large-scale X-ray facilities. In addition to measure the efficiency, operating wavelength, and focal length of a lens, we show that extremely precise measurements of the wavefront can be achieved using a laboratory source. This is obtained using our PXST method, which maps out distortions in the wavefield from the positions of features in high-magnification projection images of a scanned object. The method is well suited to the laboratory source since, like other non-interferometric wavefront sensing methods, it does not require a high degree of spatial or temporal coherence.

We demonstrate the merit of our laboratory set-up by comparing results to those obtained at a synchrotron radiation source. Using a Mo Kα source with a photon energy of 17.5 keV, we were able to prepare lenses with an operating photon energy of 60 keV, as confirmed with later measurements at that energy. Wavefront measurements obtained in the laboratory and at the synchrotron facility show excellent agreement. We estimated that we can measure the phase gradient of our optic with a precision of 18 µrad.

The knowledge of lens aberrations can be used to improve the fabrication process or produce phase correcting elements. A phase corrector was created using a 3D printer and shown to reduce the wavefront aberrations from 1.25 waves RMS to 0.19 waves RMS, indicating that our method is not only precise, but accurate. The access and availability of similar set-ups should greatly accelerate the further development of X-ray optics.

## Funding

This work was funded by DESY (Hamburg, Germany), a member of the Helmholtz Association HGF. Additional support was provided by the Deutsche Forschungsgemeinschaft (DFG, German Research Foundation)—491245950 and the Cluster of Excellence "CUI: Advanced Imaging of Matter" of DFG - EXC 2056 - project ID 390715994.

## Acknowledgements

We thank Pavel Alexeev (DESY) for the monochromator design and Manfred Spivek (DESY) for its realization. Technical support from Sabrina Bolmer, Harumi Nakatsutsumi (DESY), Luca Gelisio, Tim Gerhardt, Lars Gumprecht, Siegfried Imlau, Tjark Delmas and Julia Maracke (CFEL) is highly appreciated. Special thanks also to Wenbing Yun (Sigray, Inc.) and his team for successful collaboration and great support. We acknowledge DESY (Hamburg, Germany), a member of the Helmholtz Association HGF, for the provision of experimental facilities. Parts of this research were carried out at PETRA III synchrotron facility. We thank Olof Gutowski and Johanna Hakanpää for assistance in using beamline P07 and P11. Beamtime was allocated for proposals I-20210282 and I-20200909, respectively.

## Author Declarations

### Conflict of Interest

The authors have no conflicts to disclose.

### Data Availability

The data that support the findings of this study are available from the corresponding author upon reasonable request.